\documentclass{ws-p8-50x6-00}

\begin{document}

\title{Quark-Hadron Duality: \\ Resonances and the Onset of Scaling}

\author{W. Melnitchouk}

\address{Special Research Centre for the Subatomic Structure of Matter,\\
	Adelaide University, 5005, Australia, and		\\
	Jefferson Lab, 12000 Jefferson Avenue,
	Newport News, VA 23606, USA}

\maketitle

\abstracts{
We examine the origin of Bloom-Gilman duality and the relationship between
resonances and scaling in deep-inelastic scattering.
A simple quantum mechanical model is used to illustrate the essential
features of Bloom-Gilman duality at low momentum transfer.
As an application of local duality, we discuss model-independent relations
between structure functions at $x \sim 1$ and elastic electromagnetic form
factors.}

\section{Introduction}

Quark-hadron duality addresses some of the most fundamental issues in
strong interaction physics, providing a focus on the nature of the
transition from the perturbative to nonperturbative regions of QCD.
In its broadest form, it postulates that calculations of physical
quantities performed in either a partonic (quark-gluon) or hadronic basis
should yield identical results.
Although duality is in practice almost never realized exactly, there are
rare cases where the average of hadronic observables can be described to
good accuracy within a low order perturbative QCD treatment.

The correspondence between hadronic and partonic descriptions has been
studied in a number of physical processes, such as $e^+ e^-$ annihilation
into hadrons\cite{PQW}, heavy meson decays\cite{HQ}, and inclusive
electron-hadron scattering.
It also forms the foundation on which the theoretical framework of QCD sum
rules is based.\cite{QCDSR}
In this review, we shall discuss some recent progress made in unraveling
the origin of duality in inclusive electron scattering, first observed
some 30 years ago by Bloom and Gilman.\cite{BG}
Before discussing Bloom-Gilman duality in more detail, however, it will
be useful to briefly review the concept of duality as it first appeared
in the context of hadronic reactions.

\section{Duality in Hadron-Hadron Scattering}

Historically, duality in strong interaction physics was originally
formulated for hadron-hadron scattering, where it represented the
relationship between the $s$ and $t$ channel behaviors of scattering
amplitudes.
At low energies, the scattering amplitude, ${\cal A}(s,t)$, could be
well described by a sum over a few $s$-channel resonances,
${\cal A}(s,t) = \sum_{\rm res} {\cal A}_{\rm res}(s,t)$.
At large $s$, in the region of densely packed and overlapping resonances,
a $t$-channel partial wave series was more useful, and the high energy
behavior of ${\cal A}(s,t)$ could be described in terms of $t$-channel
Regge poles and cuts, leading to the well-known linear Regge trajectories,
${\cal A}_{\rm {I\!R}}(s,t)\ \sim\ s^{\alpha(t)}$, where
$\alpha(t) = \alpha(0) + \alpha' t$.
The puzzle confronting hadron physicists of the 1960s was how to merge
these descriptions, especially at intermediate $s$.

In analyzing $\pi N$ scattering amplitudes in the framework of finite
energy sum rules, Dolen, Horn \& Schmid observed\cite{DHS} an equivalence
or duality between $s$-channel resonances and $t$-channel Regge poles,
${\rm {I\!R}}$:
\begin{eqnarray}
\label{dualAst}
\sum_{\rm res}   {\cal A}_{\rm res}(s,t)\
&\approx&
\sum_{\rm {I\!R}} {\cal A}_{\rm {I\!R}}(s,t)\ .
\end{eqnarray}
Furthermore, as $s \rightarrow \infty$, a ``local'' version of the duality
relation (\ref{dualAst}) was found to hold,
${\cal A}_{\rm res}(s,t)\ \approx\ {\cal A}_{\rm {I\!R}}(s,t)$.

A model which unified the low and high $s$ behaviors 
was subsequently presented by Veneziano\cite{VEN}, in which the 
scattering amplitude was found to behave as
\begin{eqnarray}
{\cal A}(s,t)
&\sim& { \Gamma\left(-\alpha(s)\right) \Gamma\left(-\alpha(t)\right)
         \over \Gamma\left(-\alpha(s)-\alpha(t)\right) }\ ,
\end{eqnarray}
with poles at integer values of $\alpha(s)$ and $\alpha(t)$.
At high energy the model therefore reproduced the correct Regge behavior,
$ {\cal A}(s \rightarrow \infty, t)
\sim \Gamma\left(-\alpha(t)\right)\ (\alpha' s)^{\alpha(t)} $.

A generalization of the $s$ and $t$ channel duality, suggested by
Harari\cite{HARARI}, included both resonant and nonresonant background
contributions to cross sections.
In this ``two-component duality'', resonances were said to be dual to
nondiffractive Regge pole exchanges, while the nonresonant background was
dual to Pomeron exchange:
\begin{eqnarray}
{\cal A}
&=& \sum_{\rm res} {\cal A}_{\rm res}\ +\ {\cal A}_{\rm bckgnd}\
 =\ \sum_{\rm {I\!R}} {\cal A}_{\rm {I\!R}}
 +\ {\cal A}_{\rm {I\!P}}\ .
\end{eqnarray}

For duality in inclusive electron scattering, observed shortly thereafter
at SLAC, this naively translates into a duality between resonances and
valence quarks (for which the small $x \sim 1/s$ behavior is given by
poles on the $J^{PC}=1^{--}$ Regge trajectory), with the background dual
to sea quarks (whose small $x$ behavior is governed by Pomeron exchange).

\newpage

\section{Bloom-Gilman Duality}

In studying inelastic electron scattering in the resonance region and the
onset of scaling, Bloom and Gilman found\cite{BG} that the inclusive $F_2$
structure function at low $W$ generally follows a global scaling curve
which describes high $W$ data, to which the resonance structure function
averages.
More recently, this behavior was confirmed in high precision measurements
of the $F_2$ structure function in the resonance region at Jefferson
Lab\cite{F2JL}.
The similarity between the averaged resonance and scaling structure
functions implies that the lowest moment of $F_2$ is approximately
independent of $Q^2$.
Furthermore, the data clearly demonstrate that duality works remarkably
well for each of the low-lying resonances to rather low values of $Q^2$
($\sim 0.5$~GeV$^2$), as Fig.~1 illustrates.

\begin{figure}[h]
\vspace*{6cm}
\includegraphics{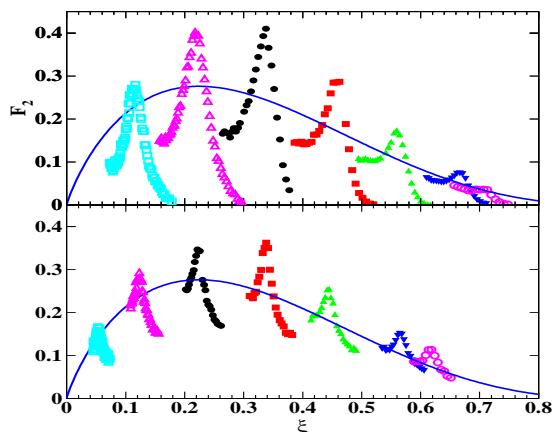}
\caption{Proton $F_2$ structure function versus the Nachtmann scaling
        variable, $\xi$, in the first (upper panel) and second (lower
        panel) resonance regions, for various $Q^2$ between 0.07~GeV$^2$
        (smallest $\xi$) and 3~GeV$^2$ (largest $\xi$), from
	Armstrong {\em et al.}\protect\cite{F2JLOC}
	The solid line is the global scaling curve determined from all
        nucleon resonance data.}
\end{figure}

Before the advent of QCD, Bloom-Gilman duality was interpreted in the
context of finite-energy sum rules\cite{DHS}, in analogy with the $s$
and $t$ channel duality observed in hadron-hadron scattering.
In QCD, Bloom-Gilman duality can be reformulated\cite{RUJ} in the language 
of the operator product expansion, in which moments of structure functions
are organized in powers of $1/Q^2$.
The leading terms are associated with free quark scattering, and are
responsible for scaling, while the $1/Q^2$ terms involve interactions
between quarks and gluons.
The weak $Q^2$ dependence of the low moments of the structure function is
then interpreted as indicating that the non-leading, $1/Q^2$-suppressed,
interaction terms do not play a major role even at low $Q^2$.

An important consequence of duality is that the strict distinction between
the resonance and deep-inelastic regions becomes entirely artificial.
To illustrate this, consider that at $Q^2 = 1$~GeV$^2$ around 2/3 of the
total deep-inelastic cross section comes from the resonance region,
$W < 2$~GeV.
However, the resonances and the deep-inelastic continuum conspire to
produce only about a 10\% correction\cite{JI} to the lowest moment of the
scaling $F_2$ structure function at the same $Q^2$.
Even though each resonance is built up from a multitude of twists, when
combined the resonances interfere in such a way that they resemble the
leading twist component.

This can be even more dramatically illustrated by considering QCD in the
large $N_c$ limit, where the hadron (or more specifically, meson) spectrum
consists of infinitely narrow, noninteracting resonances.
Since the quark level calculation still yields a smooth scaling curve,
one sees that an averaging over hadrons must be invoked even in the
scaling limit\cite{IJMV}.
In the next section we demonstrate this duality explicitly in a simple
model.

\section{Understanding the Origin of Duality}

The essential features of the dynamics behind Bloom-Gilman duality can be
exposed with the help of a simple model in which the hadronic spectrum is
dominated by infinitely narrow resonances made of valence quarks, as
discussed recently by Isgur {\em et al.}\cite{IJMV}
To strip away irrelevant details which may unnecessarily complicate the
illustration, one considers scattering from a relativistic scalar quark
confined to an infinitely massive core by an oscillator-like potential.
The virtue of such a model is that it affords exact solutions for the
complete spectrum of excited states.
Similar models have also been considered in Refs.\cite{DOM,MODELS}.
Although clearly too simple to provide a realistic description of data,
a model with these features nevertheless allows one to study the critical
questions of when and why duality holds.

If all the excited states are infinitely narrow resonances, the structure
function is given entirely by a sum of squares of transition form
factors\cite{IJMV},
\begin{equation}
{\cal S}(\nu,\vec q \, ^2)
= \sum_{N=0}^{N_{\rm max}} \, \frac{| \vec q\, |}{4 E_0 E_N}\, \,
\left| F_{0N}(\vec q\, ) \right|^2 \, \, \delta(E_N - E_0 - \nu)\, ,  
\end{equation}
where $E_0$ and $E_N$ are the energies of the ground state and $N$-th
excited state, respectively, and $F_{0N}$ is the transition form factor.
The sum over $N$ is restricted to a maximum value, $N_{\rm max}$, allowed
by the available energy transfer, $\nu$.
Note that for a scalar probe, the structure function has dimensions of
(mass)$^{-2}$.
The corresponding scaling variable,\cite{IJMV,BARB}
\begin{eqnarray}
u &=& \frac{1}{2 m} \left ( \sqrt{\nu^2 + Q^2} - \nu \right )
\left ( 1 + \sqrt{1 + \frac{4 m^2}{Q^2}} \right )\ ,
\end{eqnarray}
is defined with respect to the quark mass, $m$ (rather than the target
mass, $M$, which is infinite), and takes into account nonzero mass effects
at finite $Q^2$.
In the limit $Q^2 \to \infty$, the variable $u \to (m/M) x$.

Figure~2 shows the onset of scaling for the structure function ${\cal S}$
as a function of $u$ (the energy-dependent $\delta$-function has been
smoothed out by a Breit-Wigner shape with a given width in order to
display the $u$ dependence).
With increasing $Q^2$ the resonances are seen to move out towards higher
$u$, as observed in the physical spectrum in Fig.~1.
Furthermore, the area under the curves remains approximately constant,
indicating that global duality is reproduced by the model.
Remarkably, the resonance spikes at lower $Q^2$ also tend to oscillate
around the scaling curve, reminiscent of the local duality observed in
the proton $F_2$ data.

\begin{figure}[h]
\begin{center} 
\leavevmode
\epsfig{figure=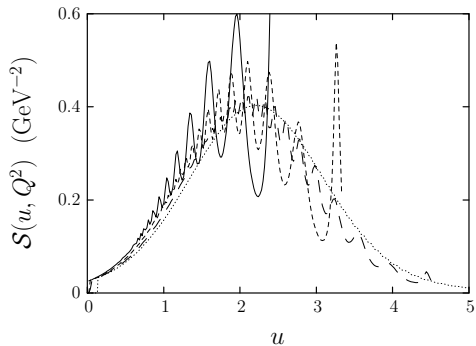,height=13cm}
\end{center}
\vspace{-9cm}
\caption{Scalar structure function versus the scaling variable $u$,
	for $Q^2=0.5$ (solid), 1 (short-dashed), 2 (long-dashed) and
	5~GeV$^2$ (dotted), from Isgur {\em et al.}\protect\cite{IJMV}}
\end{figure}

Because the curves in Fig.~2 are at fixed $Q^2$, they necessarily span
over a range of $\nu$ as $u$ varies.
As $\nu$ increases, more thresholds for creating excited states are
crossed (since $\vec q \, ^2 = Q^2 + \nu^2$), and the density of states
per $\vec q\, ^2$ interval increases correspondingly.
In fact, the correct density of states is crucial to compensate for the
falling off with $\vec q \, ^2$ of each individual form factor,
$F_{0N} \sim \vec q \, ^N \exp(- \vec q \, ^2 / 4 \, \beta^2)$,
where $\beta$ is related to the relativistic string constant\cite{IJMV}.
Had the spacing between the excited states not been Regge-like\cite{DOM},
$E_N \sim \sqrt{N}$, but linear as in a nonrelativistic solution, the
density of states would not have been sufficient to produce duality.
While it remains an interesting exercise to see whether potentials other
than a harmonic oscillator give rise to similar behavior, this simple
example illustrates that Bloom-Gilman duality appears as a natural
consequence of even the most elementary quantum mechanical systems,
with the only requirements being confinement and a correct treatment
of kinematics\cite{IJMV}.

\section{Applications of Local Duality}

If the inclusive--exclusive connection via local duality is taken
seriously, one can relate structure functions measured in the resonance
region to electromagnetic transition form factors.\cite{BG,RUJ}
Isolating an individual resonance contribution to the inclusive structure
function is problematic, however, since the separation of the resonance
from the nonresonant background is model-dependent.
For the extreme case of elastic scattering, on the other hand, there is no
background below the pion production threshold, so the extraction of the
elastic form factors from the inclusive structure function data (and vice
versa) avoids this ambiguity.

The elastic magnetic form factor of the proton has in fact been
extracted\cite{F2JL} from the recent Jefferson Lab data, and found
to agree to within 30\% with the experimental form factor for
$Q^2 < 5$~GeV$^2$.
Conversely, empirical electromagnetic form factors at large $Q^2$ can
be used to predict the $x \to 1$ behavior of deep-inelastic structure
functions\cite{BG}.
Knowledge of structure functions at large $x$ is vital for several reasons
--- the $x \to 1$ behavior is very sensitive to mechanisms for spin-flavor
SU(6) symmetry breaking, for instance, for which there are nonperturbative
and perturbative QCD predictions.\cite{MT,ISGUR}

Of particular interest is the polarization asymmetry, $A_1$, which at
large $Q^2$ is given by the ratio of spin-dependent to spin-averaged
structure functions, $A_1 \approx g_1/F_1$.
Assuming that the area under the elastic peak is the same as the area
under the scaling function (at much larger $Q^2$) when integrated from
the pion threshold to the elastic point\cite{BG}, the polarization
asymmetry at threshold can be written as\cite{EL}:
\begin{eqnarray}
\label{A1dual}
\hspace*{-0.5cm}
A_1(x_{\rm th}) &=&
\left.
\left\{ { G_M \left( G_M - G_E \right) \over 4 M^2 (1+\tau)^2 }  
+ { 1 \over 1 + \tau }
  \left( { d(G_E G_M) \over dQ^2 } + \tau { dG_M^2 \over dQ^2 }
  \right)
\right\}
\right/ { dG_M^2 \over dQ^2 }\
\end{eqnarray}
where $x_{\rm th} = Q^2 / (2 m_\pi M + m_{\pi}^2 + Q^2)$, and
$\tau = Q^2/4M^2$.
In the limit $x_{\rm th} \to 1$, corresponding to $Q^2 \to \infty$,
both $F_1$ and $g_1$ are proportional to $dG_M^2/dQ^2$, so that
$A_1^{p,n} \to 1$.
Note that SU(6) symmetry predicts that for valence quarks $A_1 = 5/9$ for
the proton and 0 for the neutron, while the perturbative QCD expectation
based on helicity conservation is $A_1^{p,n} \to 1$ as $x \to 1$.

\begin{figure}[h]
\begin{center}
\epsfig{figure=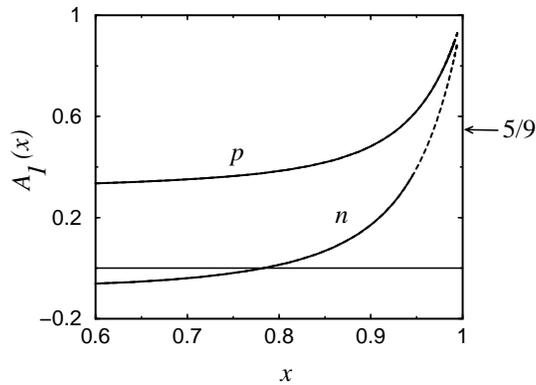,height=5cm}
\caption{Polarization asymmetries $A_1$ for the proton and neutron at
	large $x$.  The SU(6) predictions are 5/9 for $p$ and 0 for $n$.
	The dashed extensions represent asymmetries calculated from
	extrapolations of form factors beyond the currently measured
	regions of $Q^2$.}
\end{center}
\end{figure}

Using parameterizations of global form factor data, we show in Fig.~3 the
proton and neutron asymmetries $A_1$ as a function of $x = x_{\rm th}$.
The solid curves represent the asymmetry calculated from actual form
factor data, while the dashed extensions at larger $x$ illustrate the
extrapolation of the form factors beyond the currently measured regions
of $Q^2$.
Unfortunately the current data on $A_1$ extend only out to an average
$\langle x \rangle \sim 0.5$, and are inconclusive about the $x \to 1$
behavior.
While the proton $A_1$ data do indicate a rise at $x \sim 0.5-0.6$,
the neutron asymmetry is, within errors, consistent with zero over the
measured range.
It will be of great interest in future to observe whether, and at which
$x$ and $Q^2$, the $A_1$ asymmetries start to approach unity.

\begin{figure}[h]
\begin{center}
\epsfig{figure=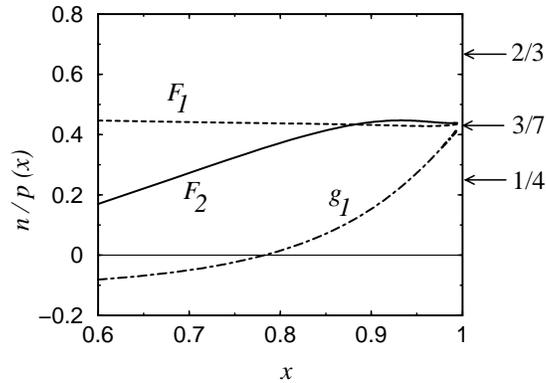,height=5cm}
\caption{Neutron to proton ratio for the $F_1$ (dashed), $F_2$ (solid)
	and $g_1$ (dot-dashed) structure functions at large $x$.
	Several leading twist model predictions for $F_2$ in the
	$x \to 1$ limit are indicated by the arrows: 2/3 from SU(6),
	3/7 from SU(6) breaking via helicity retention, and 1/4 from
	SU(6) breaking through $d$ quark suppression.}
\end{center}
\end{figure}

Expressions similar to (\ref{A1dual}) can be derived also for other
structure functions\cite{EL}.
The ratios of the neutron to proton $F_1$, $F_2$ and $g_1$ structure
functions are shown in Fig.~4 as a function of $x$, with $x$ again
evaluated at $x_{\rm th}$.
Asymptotically, each of the structure functions approaches $dG_M^2/dQ^2$,
so that in the dipole approximation the $n/p$ ratios
$\sim \mu_n^2/\mu_p^2$.
Also indicated in Fig.~4 are some leading twist model predictions for the
$F_2^n/F_2^p$, namely 2/3 from SU(6) symmetry, 3/7 from broken SU(6) with
helicity conservation, and 1/4 from broken SU(6) with scalar diquark 
dominance\cite{ISGUR}.
Note, however, that the structure functions predicted from the duality
relations contain both leading twist and higher twist contributions
(for a discussion of the conditions under which the form factors can yield
leading twist structure functions see Close \& Isgur\cite{CI}).

The reliability of the duality predictions is of course only as good as
the quality of the empirical data on the electromagnetic form factors
and resonance structure functions.
While the duality relations are expected to be progressively more
accurate with increasing $Q^2$, the difficulty in measuring form factors
at large $Q^2$ also increases.
More data on form factors at larger $Q^2$ would allow more accurate
predictions for the $x \to 1$ structure functions, and new experiments
at Jefferson Lab and elsewhere will provide valuable constraints.

\section{Conclusion}

Quark-hadron duality offers the prospect of addressing the physics of
the transition from the strong to weak coupling limits of QCD, where
neither perturbative QCD nor effective hadronic descriptions such as
chiral perturbation theory are applicable.
While considerable insight into quark-hadron duality has been gained from
recent theoretical studies, it will be important in future to understand
more quantitatively the features of the electron scattering data in the
resonance region and the phenomenological $N^*$ spectrum in terms of
realistic models of QCD.

On the experimental side, the spin and flavor dependence of duality can be
most readily accessed through semi-inclusive scattering, which requires
both high luminosity and a high duty factor.
An energy upgraded Jefferson Lab would be an ideal facility to study meson
production in the current fragmentation region at moderate $Q^2$, allowing
the onset of scaling to be tracked in the pre-asymptotic regime.
This would shed considerable light on the relationship between incoherent
(single quark) and coherent (multi-quark) processes, and on the nature of
the quark $\to$ hadron transition in QCD.

\vspace*{1cm}
{\bf Acknowledgements}

I am grateful to N.~Isgur, S.~Jeschonnek and J.W.~Van~Orden for their
collaboration on the work presented in section 4, and to experimentalists
at Jefferson Lab for stimulating discussions on duality.
This work was supported by the Australian Research Council and the U.S.
Department of Energy contract \mbox{DE-AC05-84ER40150}, under which the
Southeastern Universities Research Association (SURA) operates the
Thomas Jefferson National Accelerator Facility (Jefferson Lab).


\end{document}